\documentclass
[reviewcopy]
{elsart}

\usepackage{times}

\usepackage{graphicx}

\usepackage{amssymb}
\begin{document}

\begin{frontmatter}


\title{Conformational Transitions of Heteropolymers}

\author{Michael Bachmann\corauthref{cor1}}
\corauth[cor1]{Corresponding Author:}
\ead{bachmann@itp.uni-leipzig.de}
and
\author{Wolfhard Janke}
\address{Institut f\"ur Theoretische Physik, Universit\"at Leipzig,  \\
Augustusplatz 10/11, D-04109 Leipzig, Germany}

\begin{abstract}
We study conformational transitions of simple coarse-grained models for protein-like 
heteropolymers on the simple cubic lattice and off-lattice, respectively, by means
of multicanonical sampling algorithms. The effective hydrophobic/polar models do not require
the knowledge of the native topology for a given sequence of residues as input. Therefore
these models are eligible to investigate general properties of the tertiary folding
behaviour of such protein-like heteropolymers.     
\end{abstract}

\end{frontmatter}

\section{Introduction}
\label{secintro} 
Proteins are prominent and important examples for heteropolymers in biological systems,
where they fulfil many specific functions such as, e.g., transport of water through 
cell membranes, enzymatic activity, ATP synthase, DNA polymerization, etc. 
The specific function a protein is able to perform strongly corresponds with 
its geometric shape and this so-called native conformation is a consequence of
the folding of the chain of different amino acid residues, linked by peptide bonds, the
heteropolymer consists of. Thus the sequence of amino acids, 20 different of which
are relevant for proteins, determines the biological function of the protein. Only
a few of the possible sequences (which consist of 20 to 4000 monomers) are actually of 
importance. The main reason is that the native fold must be unique and stable, i.e., 
the conformation possesses global minimal free energy and resides in a deep, funnel-like 
valley of the free-energy landscape. It is one of the essential questions 
of protein research how the folding process towards the global energy minimum conformation, 
which takes milliseconds to seconds, proceeds and which conformational transitions
slow down the dynamics. The time scale is far too large for 
molecular dynamics simulations of realistic all-atom models with explicit solvent.
Therefore, frequently, kinetics and thermodynamic properties are studied by means of 
Monte Carlo simulations of usually extremely simplified models, which are often
based on contact matrices
of the native topology and, hence, knowledge-based and specific for the protein under
consideration. In our studies we employ more general properties of protein-like
heteropolymers. We use effective hydrophobic/polar models such
as the HP model~\cite{dill1} on the lattice and the off-lattice AB model~\cite{ab1} 
to study the influence of 
different sequences of hydrophobic and polar or hydrophilic monomers
on the strength of conformational transitions between
random coils, intermediary globules, and hydrophobic-core conformations.   
\vspace{-3mm}
\section{Conformational transitions}
\label{sectrans}
\vspace*{-3mm}
Protein structures are usually divided into four categories. The primary structure denotes the
sequence of amino acid residues. Helices, $\beta$ sheets, and hairpins are secondary
structures and the global arrangement of the monomers within a single domain is tertiary. 
Quartiary structures are formed by macromolecules with more domains. The sequence is
fixed through the genetic code and therefore single-domain proteins experience during
the folding process conformational transitions into secondary and tertiary structures.
Secondary structures are caused by the formation of hydrogen bonds. The associated 
interaction is not present in the models we studied. Our investigation
is focused on the tertiary structure, which is mainly due to the hydrophobic effect,
i.e., the formation of a core of the hydrophobic monomers separated from the 
aqueous environment by a shell of polar monomers.
\vspace*{-5mm}
\subsection*{Lattice heteropolymers}
\vspace*{-5mm}
For our studies of lattice heteropolymers we used the original
formulation of the HP model~\cite{dill1}. Two different types of monomers are distinguished:
hydrophobic ($H$) and polar/hydrophilic ($P$) residues. In the simplest form of the model
only the attractive interaction between hydrophobic monomers 
that are next neighbours on the lattice but nonadjacent along the chain is accounted for.
We performed extensive enumeration of all possible HP sequences and conformations
on the simple-cubic (s.c.)\ lattice for chains with up to 19 monomers and found a 
characteristic correspondence between 
the degeneracy of the lowest-energy states and the occurrence of a pronounced
low-temperature peak of the energetic and conformational fluctuations indicating
a transition between globular and hydrophobic-core dominated conformations~\cite{sbj1}. This means 
that lowering the temperature leads to a rearrangement of the monomers in the globules,
being globally compact conformations, and the formation of a compact hydrophobic
core surrounded by the polar monomers. 
Sequences with nondegenerate ground state (designing sequences) experience the 
strongest low-temperature transition, while it is much weaker or not present
for the other sequences. There is also another indication for a transition
between random coils and globules at higher temperature that is common to
all sequences. This behaviour is similar for longer sequences, where
sophisticated methods must be used. We developed a multicanonical
chain-growth algorithm~\cite{bj1} in order to sample lattice heteropolymers
with up to 103 monomers precisely for all temperatures~\cite{bj2}. For an exemplified
42mer with only fourfold ground-state degeneracy we found the typical
three-phase behaviour while for ten studied 48mers, whose ground states are all
high-degenerate, the low-temperature transition is rather weak.
    
It is known, however, that intermediary states are avoided in the folding 
kinetics of short realistic peptides. These miniproteins typically possess
a rather smooth free-energy landscape, where only a single barrier separates
folded and unfolded states~\cite{chan1}. Therefore, the original HP model is not sufficiently 
two-state cooperative, as the globular phase (``traps'') slows down the
folding dynamics. Simple modifications of the original HP model towards a slightly 
more realistic description seem to change, however, the folding behaviour significantly. 
A first generalization of the model is the introduction of an additional interaction between
nonbonded hydrophobic and polar monomers~\cite{tang1} based on rules derived from the
Miyazawa-Jernigan matrix of contact energies. A second modification regards the
underlying lattice type. Simulations on a face-centered cubic (f.c.c.)\ lattice are more
promising than on the s.c.\ lattice, where almost all simulations were
performed in the past, since the influence of the underlying artificial lattice is reduced. 
A prominent lattice artefact of the s.c.\ lattice, the impossibility that the $i$th and 
the $(i+2)$th monomer be next neighbours, is not present on the f.c.c.\ lattice.
In fact, with these modifications, we observed that trapping is reduced and a single peak in the
temperature dependence of fluctuations such as the specific heat indicates 
two-state folding.
\vspace*{-5mm}
\subsection*{Effective off-lattice models}    
\vspace*{-5mm}
In order to exclude artificial lattice effects we have also performed
precise multicanonical simulations of exemplified heteropolymers 
with 20 monomers using AB-like off-lattice models~\cite{bja1}. In the original formulation
of the AB model ($A$: hydrophobic, $B$: polar monomers)~\cite{ab1}, bending energy and 
pairwise residue-dependent Lennard-Jones interactions ($AA$, $BB$ contacts are long-range attractive, 
$AB$ overall repulsive) are considered.
Although there is also a general tendency to three-phase behaviour, the
intermediary states are, however, only weakly stable. Modifying the model
by introducing a curvature energy term accounting for torsion and making the
interaction of $AB$ pairs attractive at long range~\cite{ab2}, trapping is almost completely
avoided and folding proceeds in a single step.

We have also compared the folding behaviour of the heteropolymers with the
purely hydrophobic homopolymer. Although both exhibit a two-state kinetics,
the dense folds differ significantly. As expected, the homopolymer collapses
at the (finite-size!) $\Theta$ temperature from random coils to
highly compact globules. The heteropolymers, however, form a core of 
hydrophobic monomers and a polar shell. The global energy minimum conformation of the 
heteropolymers is less compact (larger mean radius of gyration),  
the more hydrophobic monomers are in the sequence.      
\vspace*{-3mm}
\section{Summary}
\vspace*{-3mm}
We have investigated conformational transitions of protein-like heteropolymers
with different simple coarse-grained lattice and off-lattice models by means
of multicanonical sampling methods. Although the identification of ``phases'',
where certain classes of conformations dominate, is difficult due to the 
impossibility of performing a scaling towards the thermodynamic limit, we
find random coil structures at high temperatures and dense conformations with
hydrophobic core at low temperatures which can be, model-dependent, separated
by intermediary globular states. The main advantage of these models for
tertiary folding is that no input, such as the sequence-dependent contact
topology of the native fold, is required. Therefore, it is possible to
study qualitatively basic principles being responsible for the cooperativity
between the phases of the heteropolymer (where the sequence of different types 
of monomers induces some kind of disorder). 

This work is partially supported by the German-Israel-Foundation (GIF) under
contract No.\ I-653-181.14/1999.
%
%
\vspace*{-3mm}
\bibliography{ccp04Bachmann}

\end{document}